**Structural Diversity in Condensed Matter: A General Characterization of Crystals, Amorphous Solids and the Structures Between**

Yueran Wang and Peter Harrowell[*]

*School of Chemistry, University of Sydney, Sydney New South Wales, 2006 Australia*

[*] Corresponding author: peter.harrowell@sydney.edu.au

Abstract

A definition of structural diversity, adapted from the biodiversity literature, is introduced to provide a general characterization of structures of condensed matter. Using the Favored Local Structure (FLS) lattice model as a testbed, the diversity measure is found to effectively filter extrinsic noise and to provide a useful differentiation between crystal and amorphous structures. We identify an interesting class of structures intermediate between crystals and glasses that are characterised by a complex combination of short-range ordering and long-range disorder. We demonstrate how the diversity can be used as an order parameter to organise various scenarios by structure change in response to increasing diversity.

## 1. Introduction

A glass is characterized by a multiplicity of local structures. Just how many and how we go about counting them is the central subject of this paper. Previously [1], we estimated that a $Cu_{50}Zr_{50}$ glass has 95 topologically distinct coordination shells. Crystals, in contrast, are made up of very few local structures. Daams and Villiers [2] reviewed over 16000 intermetallic crystals and found that 95 % of the structures consist of 4 or less distinct local coordination geometries. We shall refer to the count of the number of the local structures in a material (in whatever form this count takes), as a measure of the *local diversity* of the structure. The large gap between the crystal and glass diversities prompts the question – is there more to the structural properties of materials beyond the dichotomy of 'crystal' and 'glass'? Are there structures that lie within this gap, intermediate between the ends of the structural spectrum, or are intermediate diversities a structural 'no-man's land', unstable with respect to either the crystal or glass configuration? Is there an upper limit to the diversity in a crystal? A lower limit to the diversity of a glass? And just what types of structure exhibit these intermediate values of diversity? There are a diverse range of disordered materials that are known to retain a degree of regularity in their local order. These include complex oxides [3], gels [4], disordered molecular framework materials [5] and the melts of chalcogenide phase change materials [6]. In this paper we shall explore how an explicit measure of diversity allows us to expand our description of material structure and, as a result, allows us to develop a more comprehensive framework for exploring structural stability.

Our first task is to decide how best to identify and distinguish local structures. In 2011 Kurchan and Levine [7] proposed characterising amorphous structure based on the frequencies of different classes of configurational 'patches'. A patch is a subvolume of a configuration (typically the particles in a sphere of a given radius) [8,9] and two patches are assigned the same class if, when overlayed and rotated, their respective particles can be made to overlap to within some threshold distance. The degree of order is then quantified by the complexity defined as an entropy based on the number and populations of the classes of patches. An analysis similar to the patch classification constitutes the core of standard data compression tools like the Lempil-Ziv algorithm [10] which rely on an automated construction of a library of distinct clusters; the smaller the library needed, the greater the data compression achieved. Martiniani, Chaikin and Levine [11] have proposed using the degree of data compression as measure of structural order.

An alternate approach to cluster identification is via unsupervised machine learning based on topological descriptors (e.g. bond lengths and angles). Paret, Jack and Coslovich [12] proposed dividing particles into distinct local structural 'communities' so as to maximize the mutual information. Subsequently, these authors [13] employed dimensional reduction to identify the minimal set of descriptors needed to classify local structures. Becker et al [14] have reported the successful identification of crystalline clusters during nucleation of supercooled Zr with descriptors constructed using persistent homology.

These various autonomous strategies avoid the need to specify a target structure in advance (a feature celebrated as structural agnosticism) and, by doing so, are well suited for machine learning applications. There remains, however, something of a gap between these abstract descriptions of structure and the conventional geometric/topological approach that provides an explicit description of structure based on the distribution of different local structures present [15-17] . Some of the attractions of this latter approach are that it assists in connecting the frequency of specific structures with the stability imparted by the particle interaction potentials, allows for analysis of frustration [18], and, generally, represents a natural and intelligible extension of the conventional framework of structural analysis of materials to include amorphous solids. In this paper we develop an approach that incorporates the benefits of both approaches by adapting complexity-based measures of diversity to characterise the populations of explicitly identified local topological clusters.

This problem – quantifying the variety of distinct classes found in a population - is not unique to material science. The challenge of quantifying biodiversity in an ecological system is of central importance in ecological research and environmental monitoring [19,20] and has been the subject of significant innovation and critical assessment. In Section 2 we shall provide a brief review on the literature on measuring biodiversity and identify a number of candidate definitions of diversity. These measures will be tested on a simple model of a material, the Favored Local Structure model [21], described in Section 3, that allows considerable control over the structural diversity of the system. Using this model, we shall test our various measures of diversity in Section 4. Having selected the most reliable measure of structural diversity we will, in Sections 5 and 6, examine how diversity arises from the choice of the particle energetics and how the magnitude of the diversity determines the extended structure in the phase.

## 2. Defining Diversity: Lessons from Ecology

The literature on measuring biodiversity provides a wealth of valuable ideas and insights into the problem of quantifying complex populations [22-24]. As this literature may not be familiar to researchers in physical sciences, we provide here a brief overview.

An ecological community can be characterized by $N_i$, the number of individuals of species i. The distribution is generally represented in one of two forms; either as an abundance-rank distribution with $p_i = N_i/N_{total}$ (where $N_{total} = \sum_i N_i$) plotted against the rank i (where the i = 1 for the most frequent species, i=2 for the second most frequent species and so on) or as P(N), the number of species represented by between N and N+dN individuals. Two general strategies have been adopted to characterise these distributions. The first is to match the distribution to a model form. Three forms have proven particularly useful: a lognormal distribution [25], MacArthur's 'broken stick' distribution [26] and a geometric series [27]. Details of each distribution are provided in the Appendix.

A more general, if less complete, characterization of a species distribution is to measure an effective number of species. In 1973, Hill [28] proposed the following family of effective species number,

$$S_\alpha = \left( \sum_{i=1}^{S_0} p_i^\alpha \right)^{1/(1-\alpha)} \tag{1}$$

An associated quantity, $H_\alpha = \ln(S_\alpha)$, corresponds to a generalized entropy as proposed by Rényi [29]. (This generalized entropy has been applied to the problem of amorphous order in ref. [7].) When $\alpha = 0$, we have $S_0$, the total number of species, sometimes referred to as the 'richness' of a biological community. This measure makes no distinction between a system in which all $S_0$ species occur with equal frequency (i.e. $p_i = 1/S_0$) or one in which a single species dominates. By increasing $\alpha > 0$, we can take into account the nonuniformity of the distribution by weighting the more frequent species in the effective number. So, for $\alpha = 1$, we have

$$\begin{aligned} S_1 &= \exp\left( -\sum_{i=1}^{S_0} p_i \ln p_i \right) \\ &= \exp(H_1) \end{aligned} \tag{2}$$

where $H_1 = -\sum_{i=1}^{S} p_i \ln p_i$, is the Shannon information entropy [30]. The Shannon index, $S_1$, was used to measure diversity in our earlier study of the $Cu_{50}Zr_{50}$ glass [1]. Selecting $\alpha = 2$, we have

$$S_2 = \left( \sum_{i=1}^{S} p_i^2 \right)^{-1} \tag{3}$$

which is the reciprocal of Simpson's diversity index [31], $\sum_{i=1}^{S} p_i^2$, the probability that two members of the total population, selected at random, belong to the same species. With increasing α, we increasingly weight the significance of the more frequent species until, in the limit α→∞, we have $S_\infty = p_1^{-1}$, where $p_1$ is the relative frequency of the most common species. $S_\infty$ is known as the Berger-Parker index [32] and corresponds to an estimate of species number based on the assumption that all species occur with the maximum observed abundance. What $S_\infty$ lacks in distribution information is partially compensated for by the relative ease of estimating the abundance of the most common species. In the following analysis we shall use $S_\alpha$ to characterise the structural diversity. In Section 4 we shall select a value of α that balances an accurate representation of the number of species with the benefits of filtering out noise in the generation and sampling of configurations.

## 3. The FLS model

To explore the utility of the various measures of structural diversity described in the previous Section, we need a model material which allows for a direct adjustment of the structural diversity and can be efficiently simulated. We shall use the Favored Local Structure (FLS) model, introduced by Ronceray and Harrowell [21] in 2011, which consists of a two-dimensional triangular lattice with each site occupied by either an A particle or a B particle. (A 3D version has also been studied [33].) The local environment of a given site is defined as the pattern of A or B occupancy of the six nearest neighbor sites. There are 14 local structures (LS) that cannot be interconverted by a rotation. These distinct LS are shown in Fig. 1 along with their respective labels. Note that structures 1 to 5 are converted to structures 14 to 10 by swapping the A and B labels and that LS 7 and 8 are chiral enantiomers. We can then define a specific *system* by selecting any one or more of these distinct LS to be stable and assign them an energy of -1 (the 'favored' structures) with the remaining structures assigned an energy of zero. This assignment of energy to a configuration constitutes the many-body Hamiltonian of the model. There are ($2^{14}$-1) possible systems, each characterized by a particular choice of the set of favored local structures (FLS).

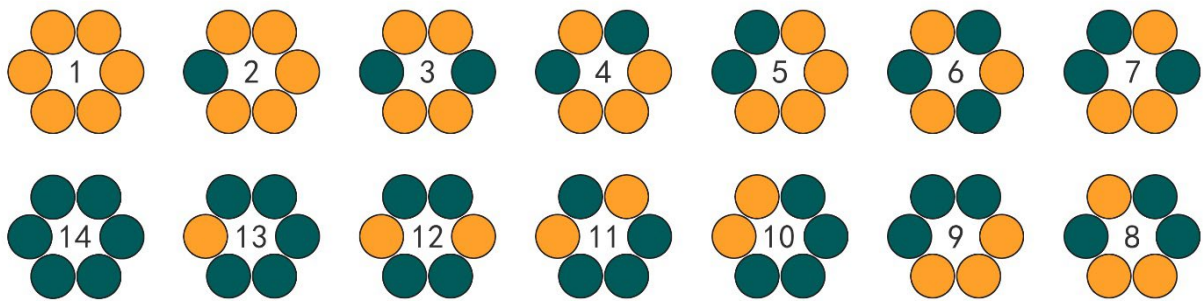

**Figure 1.** The 14 distinct local structures (LS) in the Favored Local Structure model on the 2D triangular lattice. The LS are referred to in this paper using the labels as shown in the figure.

We simulate each system at the chosen temperature using Monte Carlo (MC) sampling in which a move consists of flipping the identity of a particle from A to B or B to A so that the concentrations of A and B are not conserved. The temperature T is expressed in units of energy of a FLS. For high temperatures we have used the standard Metropolis algorithm [34], and at low temperatures, the rejection-free method due to Bortz, Kalos and Lebowitz (BKL) [35], in which sites are organized in terms of the energy change associated with the A-B exchange. The BKL method includes an estimate of the equivalent number of MC cycles (1 MC cycle = N attempted moves where N is the number of particles in the system) for each BKL move. Unless otherwise stated we have used a system size of N = 2500 and have generated groundstates by cooling from T = 1.6 at a rate of 6.25 x $10^{-5}$ per MC cycle. We have not attempted to establish whether or not the groundstates are global groundstates.

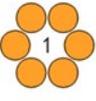

**Figure 2**. The groundstate structures for the 1FLS systems from ref. 8 and their associated number $S_0$ of local structures. For more details regarding the additional ground states of {7}, the reader is referred to ref. 36.

We have previously [21] identified the groundstate structures for the 1FLS systems. These are shown in Fig. 2. along with the associated value of $S_0$, the number of structures. We shall refer to the set of systems based on n favored local structures as nFLS systems and a specific choice of FLS will be indicated by {7}, {4,6}, {2, 9,13}, etc. Some structures have degenerate ground state structures, two in the case of {4}, {5} and {6} and at least 6 in the case of {7}. Readers will note that, with the exception of the structures associated with {1} and {14}, $S_0 > n$ (n = 1 in the 1FLS system). This means that the groundstate includes structures that are *not* favored and, thus, has an energy greater than the ideal case where FLS occupy all sites. This inability of a system to achieve the ideal minimum in energy is referred to as *geometrical frustration* and has generated a considerable literature [18,37,38], much of it focussed on the role of frustration in either supressing crystal nucleation or stabilizing amorphous phases. We shall consider the impact of frustration on structural diversity in Section 5.

One clear result of the study of the 1FLS systems, both in 2D [21] and 3D [33,39], is that they all crystallize quite readily, irrespective of the symmetry of the FLS. To achieve a situation closer to that which we find in glass forming liquids using the FLS model, we must consider systems with multiple FLS. The number of distinct systems increases rapidly with $n$, the number of FLS. We shall deal with this complexity as follows. For the $n = 2$ case we shall determine the groundstate for every possible combination of FLS's, with particular focus on the new types of structure that arise. For $n > 2$ we shall restrict ourselves to random sampling of structure combinations.

| | 1 | 2 | 3 | 4 | 5 | 6 | 7 | 8 | 9 | 10 | 11 | 12 | 13 |
|---|---|---|---|---|---|---|---|---|---|---|---|---|---|
| 2 | RC | | | | | | | | | | | | |
| 3 | S | RC | | | | | | | | | | | |
| 4 | S | S | RC | | | | | | | | | | |
| 5 | S | RC | MS | RC | | | | | | | | | |
| 6 | S | S | S | RC | C | | | | | | | | |
| 7 | S | S | S | C | C | S | | | | | | | |
| 8 | S | S | S | C | C | S | S | | | | | | |
| 9 | S | C | C | C | C | MS | C | C | | | | | |
| 10 | S | S | S | C | C | C | C | C | C | | | | |
| 11 | S | S | C | C | C | RC | C | C | C | RC | | | |
| 12 | S | S | C | RC | S | S | S | S | C | MS | RC | | |
| 13 | S | S | S | S | S | S | S | S | C | RC | S | RC | |
| 14 | S | S | S | S | S | S | S | S | S | S | S | S | RC |

**Figure 3**. The matrix of the ground states of the 91 2FLS structures where the numbers indicate a LS as indicated in Fig. 1. Crystal structures are indicated as S (a single 1FLS structure), MS (a polycrystal of two 1FLS structures), C (a crystal containing both FLS) and RC (a crystal with an intrinsically random structural component).

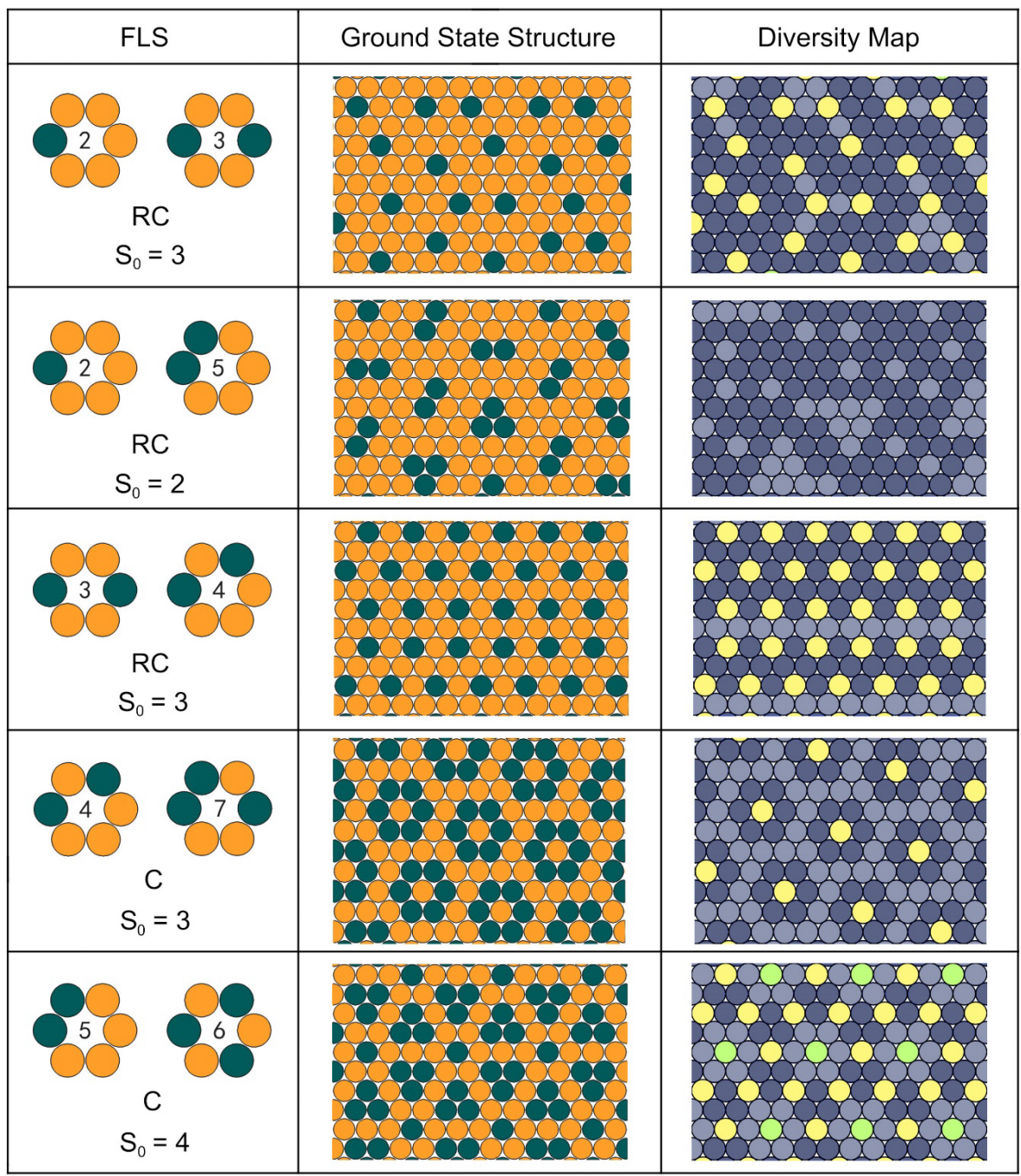

**Figure 4**. Groundstate structures for some of the 2FLS systems identified as C or RC in Fig. 3 and their associated diversity $S_0$. For each system we show the two FLS, the structure of the groundstate and the map of the LS. In this latter map, which we refer to as a *diversity map*, the FLS are indicated as dark and light grey. Non-favored structures are indicated by the yellow and green.

Our results for the structures of the groundstates for the 2FLS systems are summarised in Fig. 3. The 91 possible groundstates are designated as either a single 1FLS structure (S), a polycrystal consisting of two 1FLS structures (MS), a crystal containing both FLS (C) and a crystal containing both FLS with an intrinsically random feature (RC). Just over half of the groundstates (52) – those designated S or MS – are not new structures but stable options from the 1FLS model. These 2FLS systems has effectively 'discarded' a favored structure that was incompatible with a low energy 1FLS structure. Of the 39 structures designated C or RC, 21 correspond to distinct crystals that incorporate both FLS (the other 18 are duplicate structures generated by swapping the A and B labels). Examples of these new structures are shown in Fig. 4. The increase in the number of FLS has either resulted in the increased complexity of the crystal groundstate, as measured by the size of the unit cell, or in the appearance of intrinsic randomness in the groundstates. This randomness takes the form of the random insertion of point fluctuations, stacking faults or in the random orientation of clusters within a crystalline framework.

## 4. Diversity and Noise

As we increase the number of FLS beyond 2, the groundstate structures become both more numerous and, generally, more complex to characterize. This growing complexity, along with the growing number of different systems, makes it increasingly unfeasible to establish the true groundstate of a system to high accuracy. This means that we must deal with the extraneous structural noise introduced by the quenches. In Fig. 5 we present an example of the substantial increase in $S_0$ with increasing cooling rate. Since the kinetically trapped structures are likely to be few, we can filter some of this noise out by using a measure of diversity that weights for the more frequent LS. In Fig. 5 we show that $S_1$ and $S_2$ exhibit significantly less dependence on the cooling rate than $S_0$. and so represent attractive alternatives as a general and robust measure of the intrinsic diversity of a structural system. The trade-off is that, even in the case of perfect order (i.e. the slowest quench), $S_\alpha(\alpha > 0) < S_0$ except for the case where the order is characterised equal probabilities for all structures, in which case $S_\alpha(\alpha > 0) = S_0$ for all α.

To decide which of $S_1$ or $S_2$ is the better choice we have calculated both quantities for the 2FLS systems that are marked as C or RC in Fig. 3 and plotted, in Fig. 6, the distributions of the difference between these values and the diversity of the ideal crystals. We find that, at the standard slow cooling rate of 6.25 x $10^{-5}$ (see Fig, 6a), $S_1$ provides results closer clustered around the ideal results. At an increased cooling rate of 1.625 x $10^{-3}$, we find that $S_2$ provides the better estimate of the intrinsic diversity. On this basis, we shall use $S_1$ for our measure of diversity but note that in cases of increased structural noise (due to fast cooling rates, uncertainty in local structure identification, etc.), $S_2$ or, possibly, an even higher order $S_\alpha$, might be more useful.

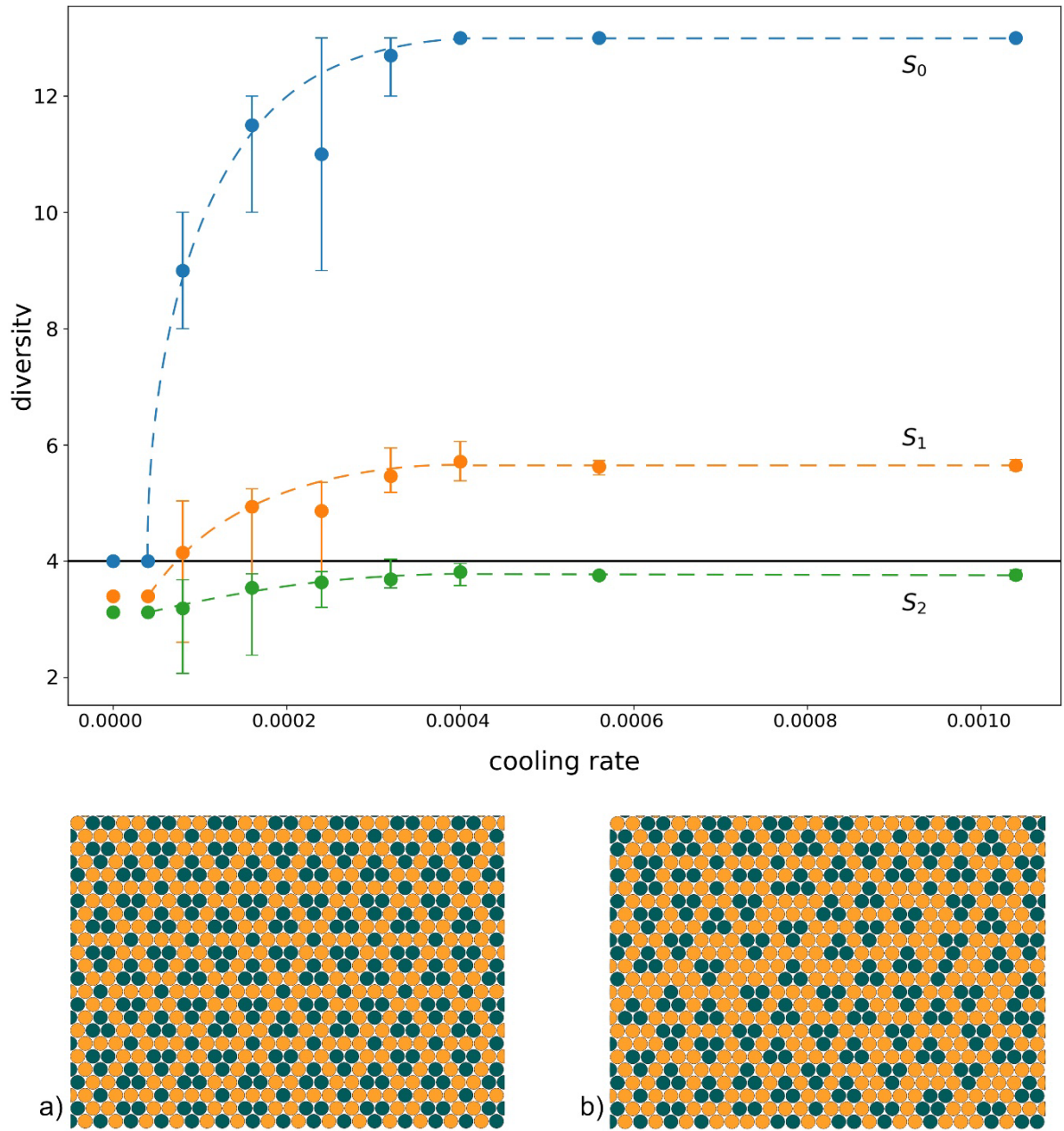

**Figure 5.** A graph showing the effect on diversity, measured using $S_0$, $S_1$ and $S_2$ for the {5,6} system, as a function of the cooling rate. Examples of the configurations from a) a slow cooling rate, $6.25 \times 10^{-5}$, and b) a fast cooling rate, $1.625 \times 10^{-3}$, are also shown. The value of $S_0$ for the perfectly ordered structure is shown as a horizontal line. The cooling rate is reported as change in reduced temperature over number of MC cycles.

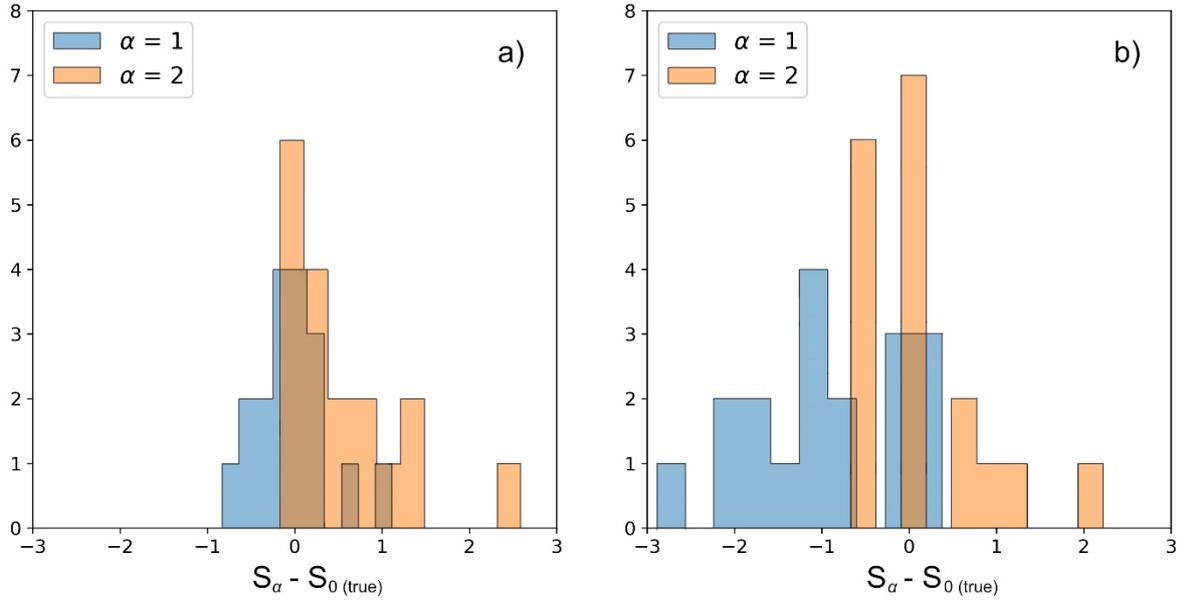

**Figure 6.** Histograms of the difference between $S_\alpha$ and the value of $S_0$ for the ideal groundstate for the 2FLS models. The choice of $\alpha = 1$ and $\alpha = 2$ are compared for a) a slow cooling rate of $6.25 \times 10^{-5}$ and b) a fast cooling rate of $1.625 \times 10^{-3}$.

In addition to fast cooling, extra structures can be introduced into a system through the influence of the choice of the dimensions of the sample. This influence of system size and shape is most marked when the groundstate is crystalline. In Fig. 7 we present the case of structure {9} where, by varying system size, we can introduce stacking faults or grain boundaries. These extended defects can increase $S_0$, significantly in some cases (as shown in Fig. 7). $S_1$, in comparison, shows little sensitivity to these artefacts.

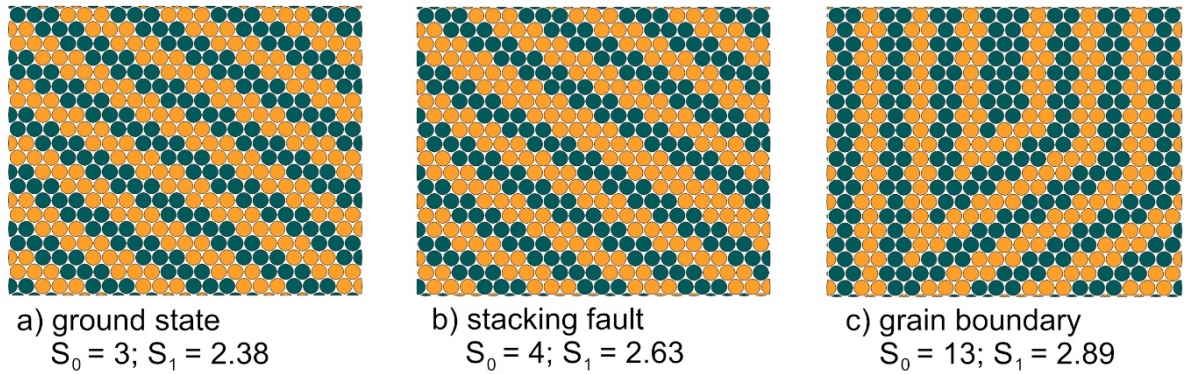

a) ground state
$S_0 = 3$; $S_1 = 2.38$

b) stacking fault
$S_0 = 4$; $S_1 = 2.63$

c) grain boundary
$S_0 = 13$; $S_1 = 2.89$

**Figure 7.** Plots of the [9] structure – the ideal structure and two examples of the structure with grain boundaries forced by an incommensurate choice of periodic boundaries. For each structure, the values of $S_0$ and $S_1$ are provided.

## 5. Diversity and the Number of Favored Local Structures

Having settled on $S_1$ as a measure of choice for structural diversity, we can now address the question of the relationship between the Hamiltonian ( in this case, the choice of FLS) and the resulting structural diversity. In Fig. 8 we have plotted the diversity $S_1$ against the number n of FLS where n is varied from 1 to 14, the entire range of possible values. As already discussed, we have only included a random sampling of the possible combination of FLS for $3 \leq n \leq 12$. Structures with energies within a threshold ($E < -0.97$) of the absolute minimum, $E = -1$, are indicated as black circles while the higher energy structures are plotted as red stars.

We find that the value of $S_1$ can differ considerably from the value of n. We find i) a positive deviation of $S_1$ from *n* that is maximum at *n=2* and decreases quickly for *n > 2* , and ii) a negative deviation of $S_1$ from *n* for larger values of n with a broad maximum around *n = 8*. We find two competing effects contributing to the diversity of the groundstates. At low values of n, the optimised groundstate includes non-favored structures. These additional structures, which we shall call *companion structures*, are a direct consequence of geometric frustration, as evidenced by their higher energy. If the FLS are unable to completely fill the space due to geometrical incompatibility, then additional structures are inevitably required to occupy these unfilled spaces. Companion structures are, therefore, a general feature of the groundstates of frustrated systems and contribute to an increase in diversity.

There is an interesting parallel between companion structures in materials and traits of biological organisms christened 'spandrels' by Gould and Lewontin [40]. Both are characterised as features that, while not intrinsically favored, occur with an elevated frequency by virtue of filling a structural niche. The purpose of the authors of ref. [40] was to challenge the assumption that a high frequency of occurrence of a feature was direct proof of its inherent favored status. In the same spirit, we note that the observed abundance of a local structure in low energy configurations cannot be automatically interpreted as evidence of that it corresponds to a low energy structure. As the number of FLS increase, structural niches will, increasingly, be filled by FLS's instead of companion structures so that frustration is eliminated with increasing n. In Fig. 8 the frustrated structures are restricted to those plotted as red stars.

The second influence on the diversity of the groundstates is the 'discarding' of FLS during the generation of the groundstates. Over a broad range of $n$, the diversity of groundstates is less than the number of favored local structures. The elimination of some favored structures arises because they cannot be accommodated in the groundstate. This behavior is well understood when the groundstate is crystalline. In our study of the 2FLS crystal groundstates we find (as shown in Fig. 3) that roughly 54% of the systems have formed a groundstate with only one FLS present. The elimination of some FLS can be obscured in the overall diversity by the addition of companion structures. The discarding of low energy structures during cooling represents clear evidence of a cooperative structural selection process. What is intriguing about the results in Fig.8 is that we see evidence of this cooperative structural selection persisting even when the groundstates have a high diversity and exhibit no obvious crystal order (as discussed in the following Section).

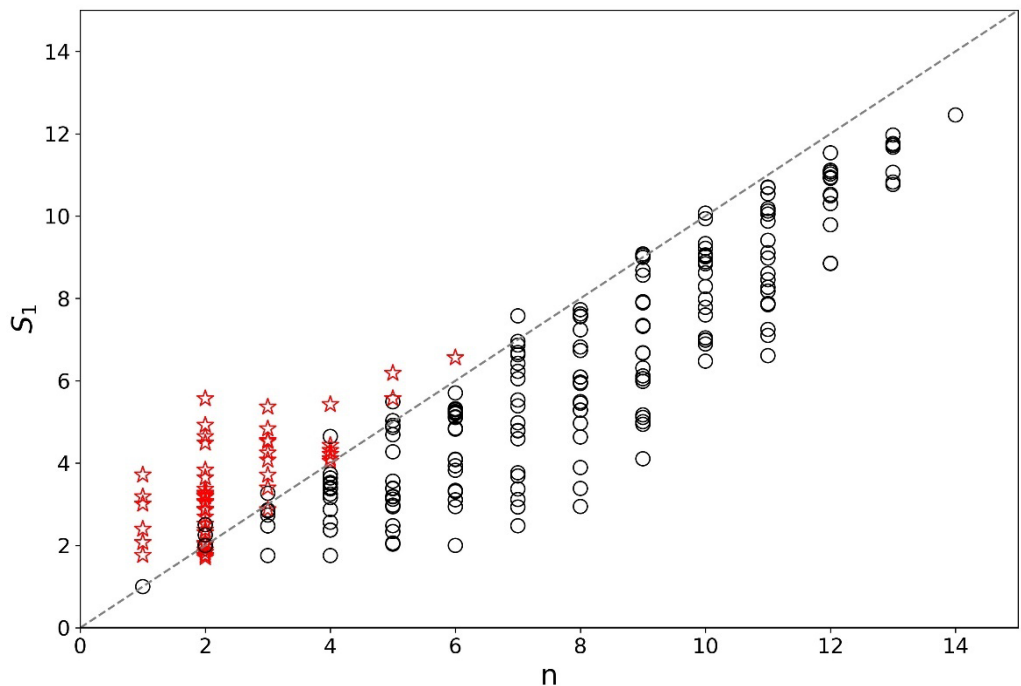

**Figure 8.** A scatter plot of the diversity $S_l$ plotted against the number $n$ of favored local structures. Structures with an energy per particle E < -0.97 are plotted as black circles. The higher energy (i.e. frustrated) structures are plotted as red stars. Note that for 2< n < 13 we have included only a randomly selected sample of the systems. The dashed line represents that case where $S_1$ = n.

## 6. Diversity and Order

We arrive, finally, at the core of our proposition that diversity is a useful approach to characterizing complex structures. In the Introduction we noted that crystals have a small number of local structures while glasses have a large number. So, can $S_1$ be interpreted as a measure of order or disorder in a material? To address this question, we need a second order parameter, one that can differentiate different degrees of ordering without requiring us to specify the details of this organization in advance. To this end we shall look at the spatial association of local structures over *intermediate* length scales. The idea is that crystal, even if composed of multiple local structures, will be characterized by only a small number of groupings of these local structures.

Consider a map of a state configuration in which each site is labelled by its local structure (i.e. a label from 1 to 14), i.e. a *diversity map* (see Fig. 4 for examples). The spatial distribution of the LS in such a map can be quantified in a manner similar to our analysis of the structure itself, i.e, by calculating the effective number of local arrangements in the diversity map. The effective number of these structures that extend out to second neighbors will be referred to as the *association number*. Each site is classified in terms of the topological arrangement of local structures of its 6 neighbouring sites. The result is a distribution of the frequencies $f_\alpha$ of these distinct non-local structural states. (Details of the definition of $f_\alpha$ are provided in the Appendix). We can then define the association number $\Sigma_1$ in a manner analogous to local diversity, i.e.

$$\Sigma_1 = \exp\left(-\sum_j f_j \ln f_j\right) \qquad (4)$$

Does $\Sigma_1$ differentiate structures with different degrees of order? We find that $\Sigma_1$ exhibits a continuous variation from 1 to ~ 1900 corresponding to structures that vary from simple crystals to completely random distributions of A and B particles. In describing the spectrum of structures it is useful to divide them, somewhat arbitrarily, into three classes. We find that groundstates characterized by values of $\Sigma_1 \leq 10$ can be reasonably identified as 'simple crystals' – some examples are shown in Fig.9a. Those configurations for which $\Sigma_1 \geq 200$ correspond to structures with little in the way of discernible structure that we shall characterise as 'amorphous'. That leaves the space between, i.e. configurations for which $10 < \Sigma_1 < 200$, examples of which have been plotted in Fig.10. These structures exhibit a wide variety of complex structures – some resembling crystals with large unit cells and complex defects and others, best characterised as random networks of repeated modular units. This collection of configurations, each characterised by evidence of structural selection at a local level and disorder over longer length scales, shall be classified as *complex order*. This classification is fuzzy. We introduce it here, not as the last word in structural classification, but as a first suggestion as to how we might differentiate the structures of disordered materials. The proposed threshold values of $\Sigma_1$ do provide us with a simple explicit classification that captures the *general* trends in the relationship between order and diversity, even if the classification of borderline structures will vary with different choices of threshold. A more objective differentiation might be possible using an extension of patch repetition analysis [8,9] where one can quantify both the maximum size of repeating patterns along with the diversity of these patterns. Complex order might correspond to structures with finite sized patch that, individually, exhibit lower diversity than the system as a whole. We leave these ideas for future work.

a) $\Sigma_1 \leq 10$

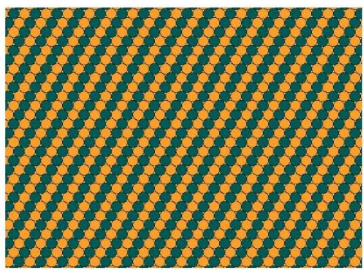 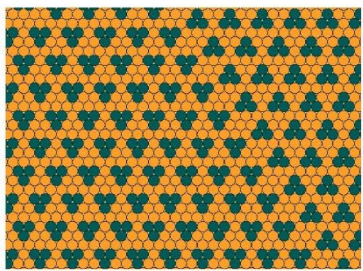 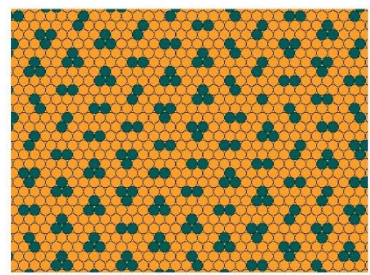

$\Sigma_1 = 2$; $S_1 = 2$
{3,12}

$\Sigma_1 = 4$; $S_1 = 2.4$
{3,5}

$\Sigma_1 = 9$; $S_1 = 1.9$
{2,5}

b) $\Sigma_1 \geq 200$

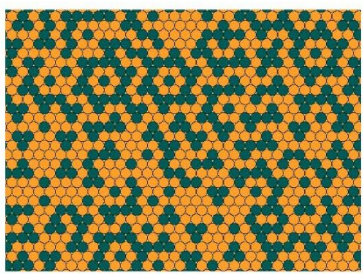 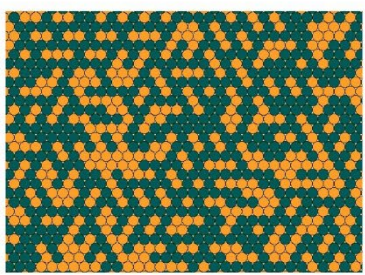 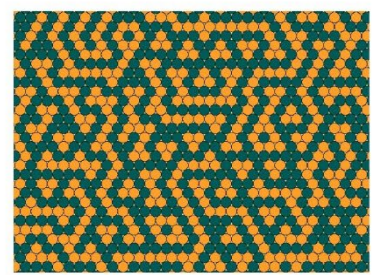

$\Sigma_1 = 257$; $S_1 = 6.2$
{1,4,5,8,11}

$\Sigma_1 = 573$; $S_1 = 6.6$
{2,7,8,9,12,13}

$\Sigma_1 = 721$; $S_1 = 7.6$
{3,4,7,8,10,11,12}

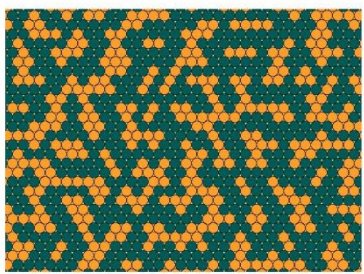 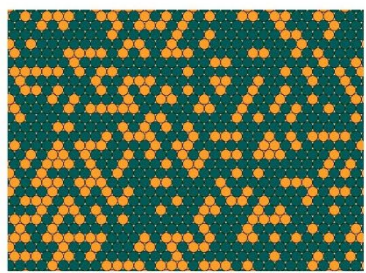 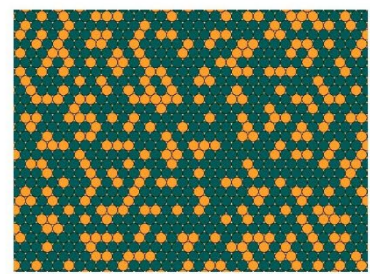

$\Sigma_1 = 1075$; $S_1 = 7.6$
{3,4,5,8,9,10,12,13}

$\Sigma_1 = 1476$; $S_1 = 8.0$
{2,3,5,7,8,10,
11,12,13,14}

$\Sigma_1 = 1849$; $S_1 = 8.8$
{1,2,3,6,7,8,9,
10,11,12,13,14}

**Figure 9.** Classifying structures using the association number $\Sigma_1$. a) Examples of configuration for which $\Sigma_1 \leq 10$. These are classified as 'simple crystals'. b) Examples of configuration for which $\Sigma_1 \geq 200$. These are classified as 'amorphous'. For each configuration we provide $\Sigma_1$, $S_1$ and the specific choice of FLS.

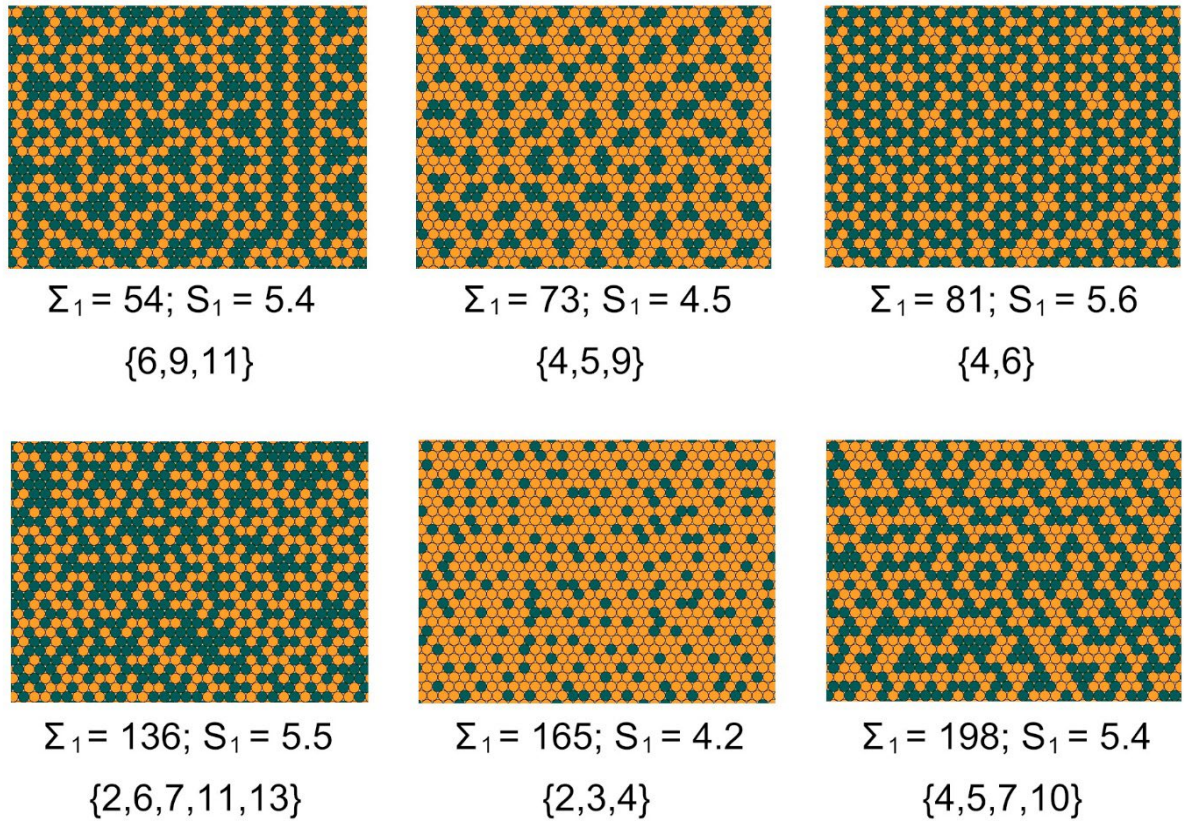

$\Sigma_1 = 54$; $S_1 = 5.4$
{6,9,11}

$\Sigma_1 = 73$; $S_1 = 4.5$
{4,5,9}

$\Sigma_1 = 81$; $S_1 = 5.6$
{4,6}

$\Sigma_1 = 136$; $S_1 = 5.5$
{2,6,7,11,13}

$\Sigma_1 = 165$; $S_1 = 4.2$
{2,3,4}

$\Sigma_1 = 198$; $S_1 = 5.4$
{4,5,7,10}

**Figure 10.** Examples of structures for which $10 < \Sigma_1 < 200$. These structures have been classified as 'complex order'. For each configuration we provide $\Sigma_1$, $S_1$ and the specific choice of FLS.

We can now address the question of the relationship between diversity and order. In Fig. 11 we present a scatter plot of the association number, plotted as $\log_{10}(\Sigma_1)$, against $S_1$ for systems generated using a variety of choices of n as indicated. We have marked out the three structural domains whose boundaries are given by $\Sigma_1 = 10$ and 200, respectively. We can make the following observations based on this data.

1) For $S_1 > 6$, we find only amorphous groundstates. Unlike amorphous states generated by kinetic arrest during fast quenches, these amorphous groundstates have energies roughly equal to that of the most stable crystal, i.e. E/N = -1. Their dominance in $S_1$>6 materials reflects the existence of multiple options for maximally stable local arrangements directly associated with the high diversity.

2) For $S_1$ > 2, we find the occurrence of groundstates exhibiting complex order. Just increasing the diversity to ~3 permits a variety of complex structures as shown in Fig. 10.

3) Simple crystal groundstates are no longer observed for $S_1 >\sim 5$ . This observation compliments the results, cited in the Introduction, regarding the apparent upper bound of local diversity in crystal structures. In the limited set of data analysed, the record for the maximum diversity exhibited by a simple crystal is held by a 4FLS structure with $S_1 \sim 4.5$ whose configuration is presented in Fig. 12a.

4) The space of structures intermediate between simple crystals and glasses - the class we have called 'complex order' - lies in the diversity range of 2 < $S_1$ < 6. Within this range, the diversity of a specific configuration is insufficient to identify the class of order that it belongs to. For example, for $4 < S_1 < 4.5$ we find examples of groundstates from all three order types.

5) We note that within the amorphous category there is a wide range of diversity, from 4 to 13. The maximum diversity corresponds to the common representation of a glass as simply a trapped configuration of a high temperature liquid. However, the existence of *low diversity* amorphous states provides us with a very different material, one in which significant cooperative structural selection is involved in reaching the disordered groundstate. The lowest diversity amorphous groundstate found so far is shown in Fig. 12b and resembles a crystal with a complex distribution of clustered point defects.

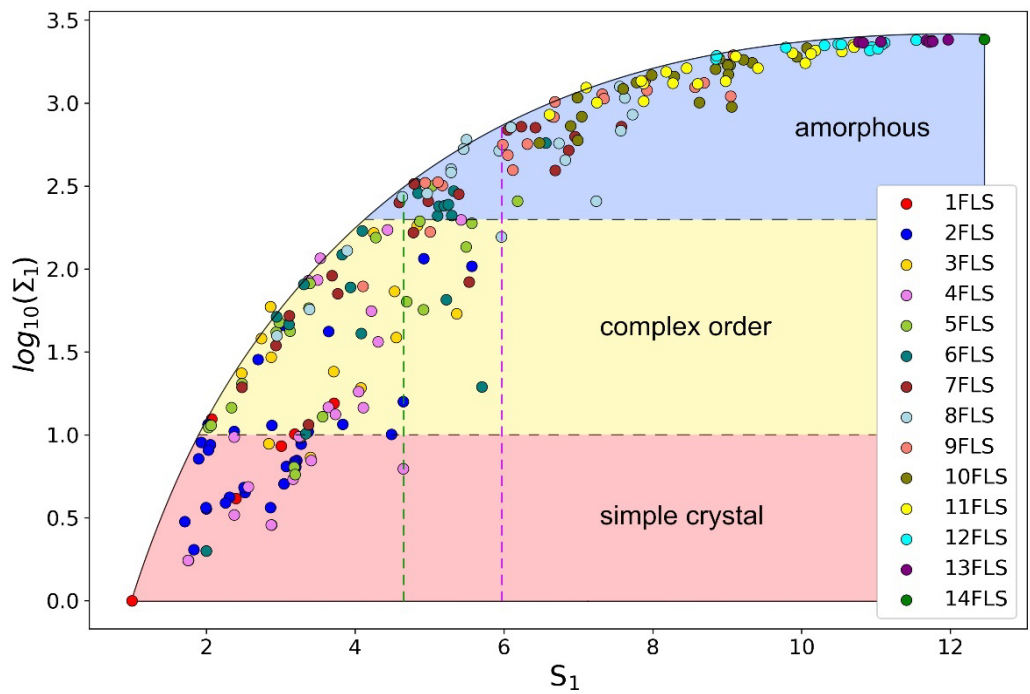

**Figure 11.** A scatter plot of $\log_{10} \Sigma_1$ against $S_1$. The three domains of order, as described in text, are indicated with boundaries at $\Sigma_1 = 10$ and $200$. The vertical dashed green line corresponds the maximum diversity for which a simple crystal has been observed. The vertical purple dashed line corresponds the value of diversity above which all structures are amorphous. The horizontal divisions between 'simple crystal', 'complex order' and 'amorphous' and the vertical lines showing diversity bounds are fuzzy as discussed in the text.

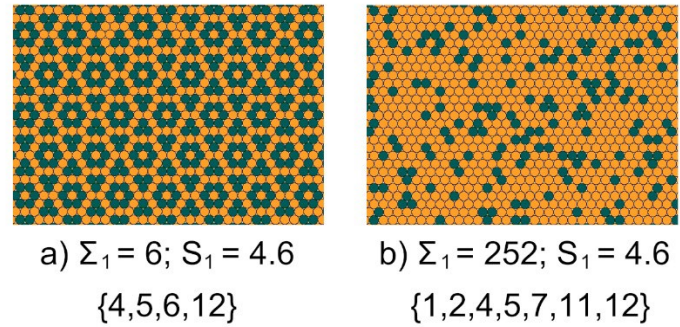

a) $\Sigma_1 = 6$; $S_1 = 4.6$         b) $\Sigma_1 = 252$; $S_1 = 4.6$

{4,5,6,12}            {1,2,4,5,7,11,12}

**Figure 12.** The configurations corresponding to a) the simple crystal with the maximum diversity and a unit cell size of 25 sites, and b) the amorphous state with the minimum diversity. Note that both configurations have the same value of $S_1$. The energy per site for structures a) and b) are E = -0.954 and -0.993, respectively. These diversity bounds depend on the fuzzy boundaries between structure types as discussed in the text.

## 7. How do Structures Incorporate Increasing Diversity: Paths from Crystal to Amorphous States

The large structural space provided by the various FLS models allow us the opportunity to organize structures as progressions, i.e. a sequence ordered by increasing diversity, from simple crystal to amorphous states. These progressions are not intended to represent any actual physical process (e.g. transformations). Instead, they provide a means of understanding how structures can vary with diversity in the form of a transect across the structural landscape. One option for this transect is the series of groundstates for a sequence of models in which we start with a 1FLS structure and then sequentially add FLS's until we get to the 14FLS model. There are 114673 such sequential progressions. The plot in Fig. 11 suggests that there are two natural choices for the progression from crystal to glass, one in which the increase in diversity follows a route of maximum $\Sigma_1$ (the 'high road') and one in which diversity increases via structures with the minimum $\Sigma_1$ (the 'low road'). In Figs. 13 and 14 we show sequences of configurations associated with these two scenarios. We find that the high road to disorder (see Fig. 13) proceeds by incorporating diversity in a crystal through the inclusion of defects. (We remind the reader these are *low energy* 'defects' constructed from favored local structures that disrupt the periodic order.) The low diversity structures correspond to the random crystals discussed in Section 3. With increasing diversity, the disorder renders the idea of 'defect' meaningless. In contrast, on the 'low road' to disorder, depicted in Fig. 14, we find that crystals incorporate diversity by initially increasing the size of the unit cell. There appears to be a threshold, around $S_1 \sim 5\text{-}6$, at which the structures cross over from crystal-like to aperiodic networks consisting of modular units.

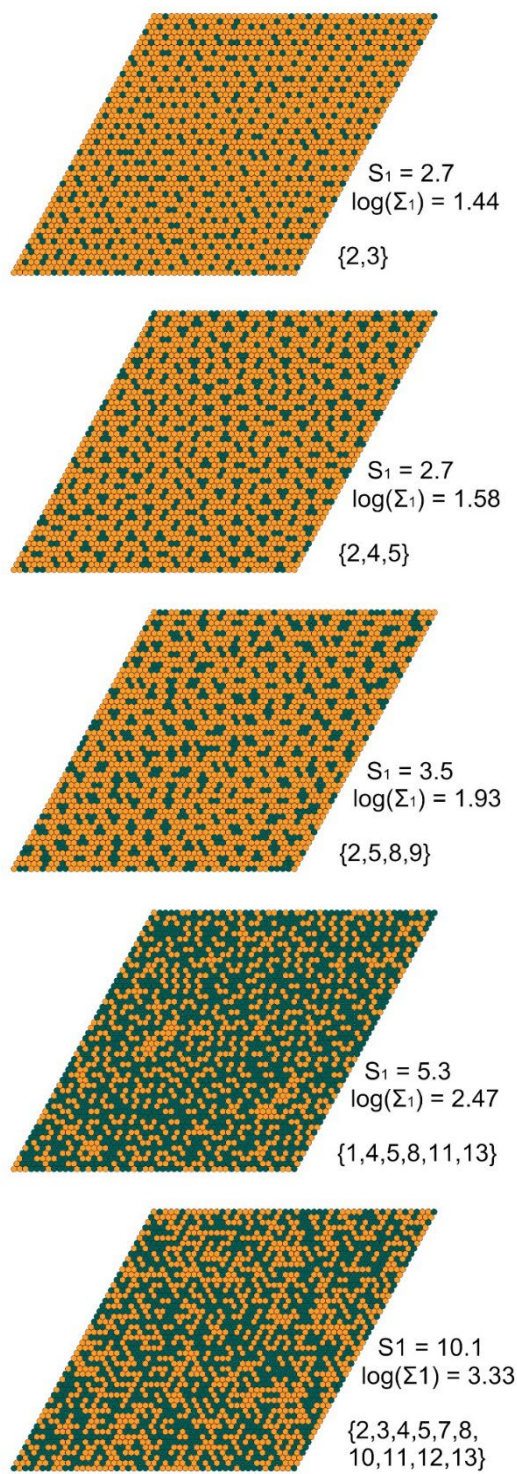

**Figure 13.** A progression of groundstates states of increasing diversity following a path of *maximum* $\Sigma_1$ (the 'high road'). The increase in diversity is accommodated by random inclusions. The values of $\log_{10} \Sigma_1$ and $S_1$ are shown along with the structures used to generate these images.

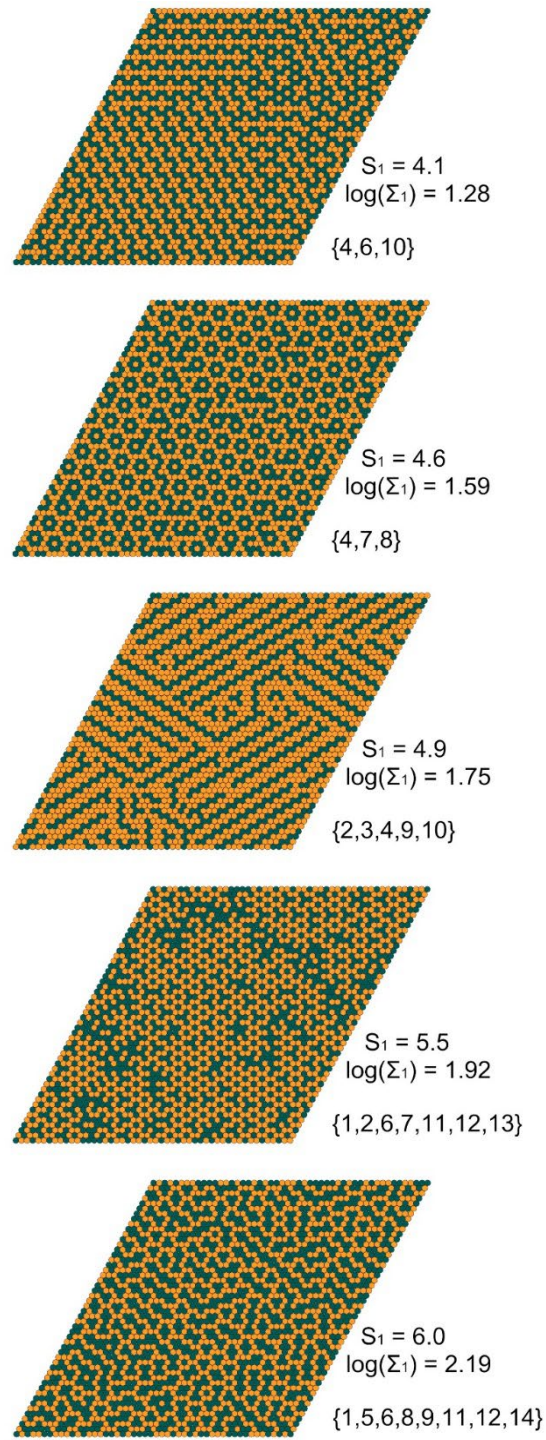

**Figure 14.** A progression of groundstates of increasing diversity following a path of *minimum* $\Sigma_1$ (the 'low road'). The increasing diversity is, initially, incorporated via expanded unit cells which, with increasing $S_1$, are supplemented by increasingly complex grain boundaries. The structures used to generate these images are shown.

## 8. Discussion

The goal of this paper is to introduce the concept of structural diversity to provide a general characterization of the structure of materials, one detached from the specifics of unit cells and crystal periodicity. In this Section we shall discuss some of the outcomes of this study.

### 8.1 On the Existence of Complex Order

One interesting use of diversity is its capacity to characterize structures that lie between the extremes of simple crystals and the kinetically-arrested glasses. As pointed out in the Introduction, the list of *possible* examples of this intermediate 'complex' order includes oxide glasses [3], gels [4], disordered molecular framework materials [5] and the melts of chalcogenide phase change materials [6]. With regards the complex order exhibited by the FLS model, two qualifications are warranted. The first is to acknowledge that this group of structures will include groundstates that, on slower cooling, are actually crystalline. Our designation 'complex order' should, therefore, be regarded as tentative, an identification of a class of structures that do exist, at least as local minima. The second qualification concerns the role of dimensionality on the possible range of intermediate structures. As discussed in Appendix Section A.2, we find that the upper bound of the association number $\Sigma_1$ increases with $S_1$ as $\sim S_1^{z/2}$, where the coordination number $z$ is the number of nearest neighbors. Using this formula, we reach the amorphous threshold ($\Sigma_1 = 200$) at $S_1 = 5.8$ for the 2D lattice (larger than the observed value of 4.4). In the case of the 3D FCC lattice with $z = 12$, however, this amorphous threshold is achieved by $S_1 = 2.4$. This rapid rise in $\Sigma_1$ with increasing $S_1$ suggests that the diversity range of possible complex order will be significantly reduced (or even vanish) in 3D. Clearly, the limitations of the FLS model leave the question of the existence of complex order in more realistic models an interesting open question.

**8.2 On the Structural Consequences of a Continuous Increase in Diversity**

In Figs. 13 and 14, we presented two scenarios of how structure might change as the diversity increases. The possibility of treating diversity as an order parameter, one capable of continuous variation, represents its particular usefulness. In considering this aspect of structural diversity, it is useful to compare this approach with defect-induced amorphization, a phenomenon that has received considerable attention and represents the simplest example of the important low temperature route to amorphous states by mechanical means. In 1992, Fecht [41] proposed a thermodynamic model for the disordering transition, analogous to thermal melting. A number of experimental studies involving either mechanical induced [42] or compositionally induced [43] defects have confirmed the first-order character of the defect induced amorphization. While the proposed diversity measure could be usefully applied to these problems, our specific results for the FLS model differs significantly from these earlier studies. The defects considered in refs. [41-43] are high energy defects that result in increasing the fictive temperature of the crystal to the point of melting. In contrast, the disorder we consider arises from low energy structures, resulting from changes in the Hamiltonian. Disordering arises not by destabilizing the crystal phase but, rather, by rendering it irrelevant amongst a growing population of disordered groundstates with the same (or lower) energy.

**9. Conclusions**

In this paper we have introduced a measure $S_1$ of the diversity of local structure with general applicability across all condensed states of matter: crystalline, polycrystalline, networks and glasses. We have demonstrated that the measure, based on the Hill numbers $\{S_\alpha\}$, can be tuned to filter out (or specifically select for) structural noise. This capacity – to characterize and differentiate structures that range from ideal crystal to the 'idealized' glass (the latter being the arrested configuration of a high energy liquid) – represents a valuable tool for categorizing structures of arbitrary complexity and variety. While we have used a lattice model to generate the different structures, our analysis can be applied directly to the Voronoi clusters of particle models as demonstrated in ref. [1].

When applied to the groundstates of the 2D FLS model, we established that the structural diversity exhibits a continuous variation across the total accessible range. Between the low diversity crystals and the high diversity amorphous solids and the minimum diversity, we identified and class of an intermediate class of structures, labelled complex order, that were characterised by a combination of strong structural correlations and significant disorder. It is an interesting question as to whether these intermediate groundstate structures have analogues for Hamiltonians based on realistic particle interactions. Finally, we organized groundstates into sequences arranged in order of increasing diversity to explore scenarios by which crystallinity evolves into amorphous order as the number of local structures increased.

This is a preliminary study to establish of a useful measure of structural diversity. Having done so we conclude with the following suggestions for new approaches to the treatment of disorder for future work. The diversity and the relative abundances of the local structures distinguish a group of inherent structures. It would be useful to determine whether this distinguishing diversity persisted as the temperature is increased. If so, it would allow us to monitor the role of different groups of inherent structures in the structural fluctuations of supercooled liquids. There is a considerable literature in ecology in the relation between diversity and stability [44-46]. Whether analogous instabilities between diversity-defined groups of groundstate structures are a feature of structural fluctuations in dense disordered phases is an intriguing possibility.

**Acknowledgements**

PH would like to acknowledge valuable conversations with Toby Hudson and the timely introduction to the subject of ecology by Lee Dyer, Christopher Jeffery and Craig Dodson. YW gratefully acknowledges support in the form of an Australian Postgraduate Award from the Australian Federal Government.

**Dara Availability Statement**

The data that support the findings of this study are available from the corresponding author upon reasonable request.

**Appendix**

**A1. Commonly Used Functions to Describe Species Distributions**

The following three distributions represent potentially useful approaches to characterising and classifying different distributions of local structure. Each distribution has proven useful in describing observed species distributions. The three distributions are:

i) the lognormal distribution, given by

$$S(N) = S_o \exp\left(-\left(\ln N - \ln N_o\right)^2 / 2\sigma^2\right) \tag{A1}$$

This distribution was found to match species distributions from a variety of studies by Preston in 1948 [25].

ii) MacArthur's 'broken stick' distribution [26]

$$N_i = \frac{N_T}{S_0} \sum_{n=i}^{S_0} \frac{1}{n} \tag{A2}$$

where $i$ is the rank and $S_0$ is the total number of species. For $S_0 >> 1$, $N_i \simeq \left( N_T / S_0 \right) \ln \left( S_0 / i \right)$. This distribution is related to the case where a fixed number $N_T$ of individuals are randomly distributed among $S_0$ species ,

iii) a geometric series [27] takes the form

$$N_i = N_T C_k k (1-k)^{i-1} \ \text{where} \ C_k = \left( \sum_{i=1}^{S_T} k(1-k)^{i-1} \right)^{-1} = \left( 1 - (1-k)^{S_T} \right)^{-1} \qquad \text{(A3)}$$

We arrive at this distribution when $N_1$, the number of the top ranked species, is a fraction k of $N_T$ and $N_2$ is equal to k times the remaining number of individuals and so on.

## A2. Calculation of the Extended Diversity

In the form of the diversity map, the local structures are considered as two digit numbers, i.e. '01', '02' … '14'. For every given site, the LS of the six neighboring sites are connected as a twelve digit string, which refers to the code of a unique topological arrangement. Due to the large possibilities of arrangements in the diversity map, the difference in LS order and their mirror states are not considered distinct arrangements and are classified under the same code. The frequency $f_a$ is then defined as

$f_a$ = (number of sites classified under code $a$) / (total number of sites).      (A4)

The dependence of $\log_{10} \Sigma_1$ on the diversity $S_1$ can be characterized by two bounding curves as shown in Fig. 15. The upper bound, in which $\Sigma_1 \sim S_1^{3.6}$, can be rationalized as the random filling of roughly half the nearest neighbour sites by local structures drawn from $S_1$ structures present. (The factor of ~ ½ accounts for the constraints associated with the overlap of neighbouring local structures.)  We can generalize this upper bound to other lattices with

$$\Sigma_1 \sim S_1^{z/2} \qquad \text{(A5)}$$

where z is the coordination number.

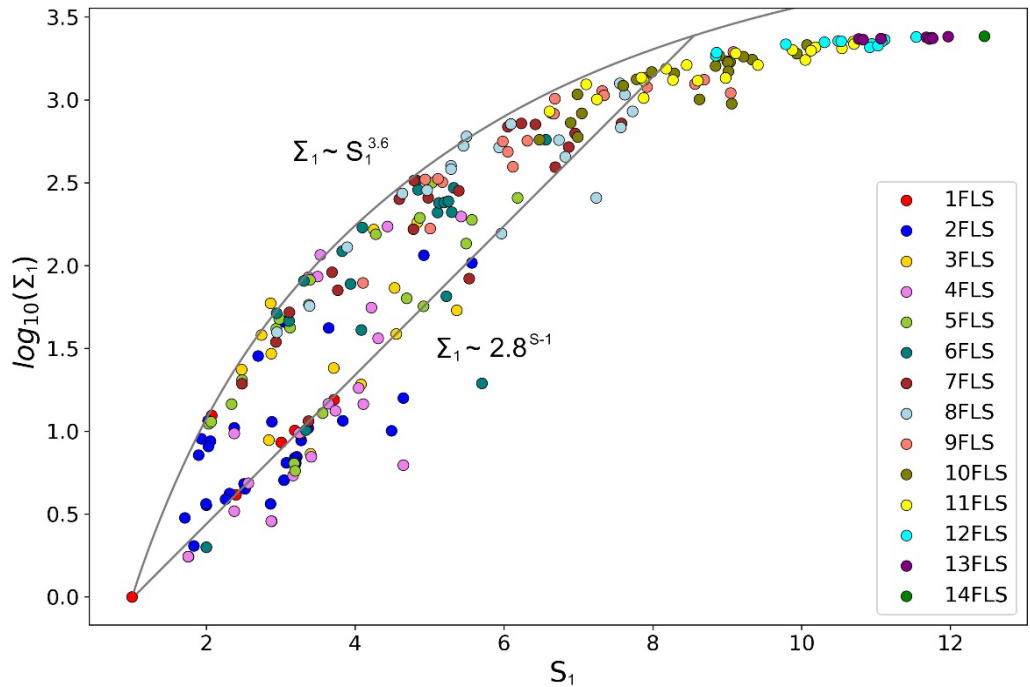

**Figure 15.** A plot of $\log_{10}\Sigma_1$ vs $S_1$ that includes fitted functions, as indicated, for the upper and lower bounds of $\Sigma_1$ as a function of $S_1$.